\documentclass[twocolumn, preprintnumbers, superscriptaddress,
  nofootinbib]{revtex4-1}
\pdfoutput=1
\usepackage[pdftex]{graphicx}
\usepackage{amssymb,amsmath}
\usepackage[colorlinks,linktocpage]{hyperref}

\begin{document}
\title{Scattering of linear waves on a soliton}
\author{A.S. Dmitriev}\email{dmitriev.as15@physics.msu.ru}
\affiliation{Institute for Nuclear Research of the Russian Academy
  of Sciences, Moscow 117312, Russia}
\author{E.A. Dmitrieva}\email{kate\_steklova@mail.ru}
\affiliation{Lomonosov Moscow State University, Faculty of Physics,
  Moscow 119991, Russia}
\affiliation{Institute for Nuclear Research of the Russian Academy
  of Sciences, Moscow 117312, Russia}
\author{A.G. Panin}\email{panin@ms2.inr.ac.ru}
\affiliation{Institute for Nuclear Research of the Russian Academy
  of Sciences, Moscow 117312, Russia}

\begin{abstract}
  We have shown that the wave scattering by a soliton occurs in a peculiar
  way. The nonlinear interaction leads to the generation of waves with
  frequencies that are multiples of the frequency of the incident wave, minus
  the frequency of the soliton, while the soliton experiences growth due to
  the absorption of particles from the incoming wave. We propose an analytical
  approach based on the Born approximation to calculate the amplitudes of
  scattered waves and, consequently, the number of particles absorbed by the
  soliton. This approach is validated through direct comparison with the
  results of numerical simulation.
\end{abstract}

\maketitle

\section{Introduction}
\label{sec:1}

Solitons are stable bound states which exist on a classical as well as
quantum mechanical levels in nonlinear field theories. They have a finite
stable shape in space, which, unlike wave packets formed by superpositions of
plane waves, preserves over time and during free propagation. The dispersion
in the solitons is balanced by focusing nonlinear effects. Since their first
observation on the surface of water~\cite{russell1844report}, solitons have
been widely studied both theoretically and experimentally in many branches
of physics such as optics~\cite{kivshar2003optical}, plasma
physics~\cite{Ichikawa_1979}, condensed matter physics~\cite{BISHOP19801},
cosmology~\cite{KUSENKO199846} and other natural sciences.

There are several exactly integrable nonlinear theories, in which
solitons can be obtained by means of the inverse scattering
transform~\cite{Lax1968,Zakharov1971}. These theories play an exceptional
role because they provide universal mathematical tools for studying some
general physical phenomena~\cite{novikov1984theory, drazin1989solitons}.
In particular, in exactly integrable case the interactions between colliding
soliton and wave packet or another soliton are purely
elastic~\cite{zakharov1973interaction}. The soliton recovers its exact
initial shape and velocity after a collision.

In real physical problems, however, various perturbations usually occur that
violate the exact integrability. These not only modify the soliton
properties, but also,  under certain conditions, lead to completely new
phenomena~\cite{Kishvar1989}. In particular, collisions between solitons
under influence of higher order nonlinear terms exhibit a variety of outcomes,
including fusion, fission, annihilation~\cite{Gatz1992, Tikhonenko1996,
  Nguyen2014} and emission of quasi-linear waves~\cite{Kishvar1989}. Another
example is the transmissions of solitons through linear and nonlinear
inhomogeneities which proceeds with increased rates under certain
conditions~\cite{garnier2006transmission}. 

In this paper, we consider the scattering of linear high-frequency
waves on non-topological solitons in the model of a nonlinear
non-integrable Schr\"odinger equation. We show that, in contrast to the
integrable case in which the incoming wave causes a shift in the soliton's
position~\cite{haus1996interaction}, the presence of non-integrable nonlinear
interaction leads to a non-trivial scattering picture. Nonlinear interactions
give rise to waves of double, triple, and so on, frequencies, while the
soliton grows by absorbing particles from the incoming wave. We propose
an analytical method for describing the scattering pattern of high-frequency
waves on a soliton. To validate the method, we compare its predictions with
the results of numerical simulations and observe a strong agreement.

The paper is organized as follows. In Section~\ref{setup} we present the
general setup and introduce the notations that will be used throughout
the paper. An explicit numerical example of scattering of a wide Gaussian
wave packet on a soliton is considered in Section~\ref{numerics}.
In Section~\ref{pert} we address the computation of the amplitudes
of the waves transmitted through and reflected from the soliton, and
compare analytic predictions with the data of numerical simulations.
In Section~\ref{discus} we discuss the results. Appendix~\ref{numer}
includes details of method used in numerical simulations.

\section{Setup}
\label{setup}
We start by considering the nonlinear Schr\"odinger equation in
dimensionless variables
\begin{equation}
  \label{schrod}
  i \partial_t \psi = -\frac{1}{2}\partial^2_x\psi+V(|\psi|^2)\psi\;,
\end{equation}
where the last term introduces nonlinearity.
Evolution governed by Eq.~\eqref{schrod} conserves a number of quantities:
the particle number (norm) $N$,
\begin{equation}
  \label{N}
  N = \int dx\, |\psi|^2\;,
\end{equation}
the energy of the system $E$,
\begin{equation}
  \label{E}
  E = \int dx\,\left [ \frac{1}{2} |\partial_x\psi|^2
    + \int\limits_0^{|\psi|^2} ds\, V(s) \right ]\;,
\end{equation}
and the total momentum $P$,
\begin{equation}
  \label{P}
  P = i \int dx\, \left [ \psi \partial_x\psi^*
    - \psi^* \partial_x \psi \right]\;.
\end{equation}

Suppose that Eq.~\eqref{schrod} admits a soliton solution of the form
\begin{equation}
  \label{sol}
\psi_s(t,x) = f(x)\mathrm{e}^{-i\gamma t}\;, 
\end{equation}
where real function $f(x)$ gives the soliton profile. It satisfies
$f(-x) = f(x)$ and $f(x) \to 0$ for $|x| \to \infty$. The soliton frequency
is negative,  $\gamma < 0$. It determines the binding energy of nonrelativistic
particles in a soliton according to the relation
\begin{equation}
  \label{dEsol}
  dE = \gamma dN\;,
\end{equation}
which directly follows from Eqs.~\eqref{schrod},~\eqref{sol}.
The solution is supposed to be classically stable in according to the
Vakhitov-Kolokolov stability criterion~\cite{vakhitov1973}.

Below we will consider a wave packet moving from large negative $x$ to the
right and scattering on a soliton~\eqref{sol} centered at $x = 0$,
\begin{equation}
  \label{psi0}
  \psi_0 = A(t,x) \mathrm{e}^{-i\omega t + i p x + i\phi_0}\;,
  \qquad \omega = \frac{p^2}{2}\;,
\end{equation}
where $A$ specifies a wave packet shape and $\phi_0$ is a constant phase.
In what follows, we will assume that $|V(A^2)| \ll \omega$, so the wave packet
freely propagates to the region of interaction with the soliton. In addition,
we will consider wave packets with a width $\sigma$ much larger than the
wavelength, $\sigma \gg 2\pi/p$ so that the change in its shape during 
movement can be neglected.

\section{Numerical illustration}
\label{numerics}
To begin with, let us consider the scattering of a wave packet on a soliton
numerically. For this purpose, we select a potential
\begin{equation}
  \label{pot}
  V(|\psi|^2) = -\lambda |\psi|^2 + g |\psi|^4\;.
\end{equation}
In what follows we will use $\lambda = g = 1$. Nontrivial couplings can be
restored by the redefinitions $\psi \to \psi \,\sqrt{\lambda/g}$,
$x \to x\, \lambda/\sqrt{g}$, $t \to t\, \lambda^2/g $.

\begin{figure}[t]
  \centering
  \includegraphics{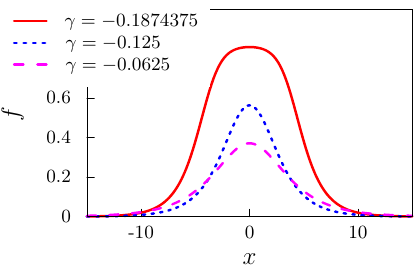}
  \caption{The soliton profile~\eqref{profile} for different frequencies.}
  \label{fig:pot}
\end{figure}


The nonlinear Sch\"odinger equation with the potential~\eqref{pot} admits
the soliton solution~\eqref{sol}, which exists for $-3/16 < \gamma < 0$. It is
remarkable that the soliton profiles are known analytically, 
\begin{equation}
  \label{profile}
  f(x) = \frac{2\sqrt{-\gamma}}
  {\left (1 + \sqrt{1 + \frac{16}{3} \gamma}\,
      \mathrm{cosh\left(2\sqrt{-2\gamma}x \right)}\right )^{\frac12}}\;.
\end{equation}
Function $f(x)$ for different $\gamma$ is plotted in the Fig.~\ref{fig:pot}.

Now, let us consider the scattering of a wide Gaussian wave packets on the
soliton. We solve numerically the nonlinear Schr\"odinger
equation~\eqref{schrod} with the potential~\eqref{pot} starting at $t = 0$
with the initial wave function $\psi = \psi_0 + \psi_s$. Here $\psi_s$ is the
wave function of the soliton centered at $x = 0$ and $\psi_0$ is an incident
Gaussian wave packet, Eq.~\eqref{psi0} with
\begin{equation}
  \label{gauss}
  A(0,x) = A\,\mathrm{e}^{-(x-x_0)^2/2\sigma^2}\;,
\end{equation}
localized at large negative $x \simeq x_0$ far away from the soliton. The
details of the numerical method are presented in Appendix A.
The results are demonstrated in Fig.~\ref{fig:evol} and in the
movie~\cite{movie}.
\begin{figure}
  \centering
  \includegraphics{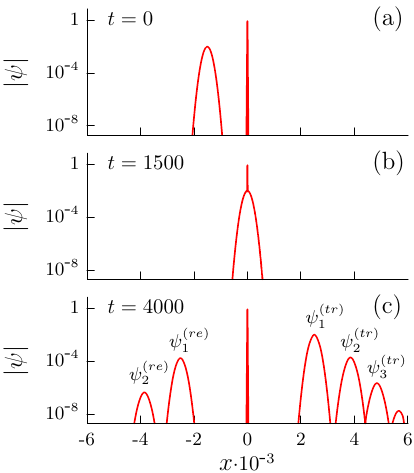}
  \caption{Results of numerical simulation of scattering of a Gaussian wave
    packet ($A = 0.01$, $x_0 = -1500$, $\sigma = 100$ and $p = 1$) on a soliton
    ($\gamma = - 0.1874375$). Frames (a)-(c) displays $|\psi|$ at different
    times.}
  \label{fig:evol}
\end{figure}
The incident wave packet strikes the soliton and then separates into a
reflected and a transmitted parts. Each part consists of several wave packets
moving with different velocities (see Fig.~\ref{fig:evol}c). In
particular, the wave packets with maximal amplitudes, transmitted
$\psi_1^{(tr)}$ and reflected $\psi_1^{(re)}$, move at the velocity of the
incident wave packet $v_1 = \sqrt{2\omega}$. They are the result of
quantum-mechanical-like scattering on a potential barrier produced by a
soliton. The wave packets denoted by $\psi_2$ and $\psi_3$ in Fig.~\ref{fig:evol}c have frequencies $\omega_2 = 2\omega - \gamma$,
$\omega_3 = 3\omega - \gamma$ and velocities $v_2 = \sqrt{4\omega - 2\gamma}$, $v_3 = \sqrt{6\omega - 2 \gamma}$ respectively and appear due to nonlinear
interaction of an incident wave packet and a soliton. We have numerically
verified that for $A \lesssim 0.01$ the amplitude of $\psi_1^{(re)}$ scales as
$A$, while the amplitudes of $\psi_2$ and $\psi_3$ are proportional to $A^2$
and $A^3$ respectively. This fact will be used in the next section
to construct a perturbation theory with respect to $A$.

Let us demonstrate that in the scattering picture described above a certain
number of particles from the incident wave packet get stuck in the soliton
region. Initially, the total number of particles is the sum of the number
of particles in the incident wave packet $N_1^{in}$ and in the soliton
$N_s^{in}$. After scattering the number of particles in the wave packets
$\psi^{(re)}_2$ and $\psi_3^{(tr)}$ is exponentially small and can be
neglected, see Fig.~\ref{fig:evol}c. Thus, we have
\begin{equation}
  \label{Nlaw}
  N_1 + N_2^{(tr)} + N_s =  N_1^{in} + N_s^{in}\;,
\end{equation}
where $N_1$ and $N_2^{(tr)}$ are the number of particles in the wave packets
$\psi_1^{(tr)} +\psi_1^{(re)}$ and $\psi_2^{(tr)}$ respectively and $N_s$ takes
into account particles in the soliton, including localized states that can be
excited as a result of the scattering. Similarly, for the total
energy one can write
\begin{equation}
  \label{Elaw}
  E_1 + E_2^{(tr)} + E_s =  E^{in}_1 + E^{in}_s\;,
\end{equation}
where $E_s$ also includes the kinetic energy of the soliton after scattering
and the energy of the localized modes. The energy of the wave packets,
considering that $|V(A^2)| \ll \omega$, equals to the product of the frequency
by the number of particles, $E = \omega N$. Then from
Eqs.~\eqref{Nlaw},~\eqref{Elaw} we can obtain
\begin{equation}
  \label{dE}
  (\omega - \gamma)\Delta N_1 + (2\omega - \gamma)\Delta N_s
  = \Delta E_s\;,
\end{equation}
where $\Delta N_{1,s} = N_{1,s} - N^{in}_{1,s}$ and $\Delta E_s = E_s-E^{in}_s$.
Let us show that Eq.~\eqref{dE} has no solution with $\Delta N_s = 0$.
Indeed, in this case the number of particles in the wave packets $\psi_1$
decreases by $N_2^{(tr)}$ after scattering (see Eq.~\eqref{Nlaw}), thus the
first term in the right-hand side of Eq.~\eqref{dE} is negative. On the other
hand, a soliton with constant number of particles can only increase its
energy, for example, by acquiring kinetic energy. Thus one concludes that
$\Delta E_s$ must be positive, which contradicts Eq.~\eqref{dE}.

\begin{figure}
  \centering
  \includegraphics{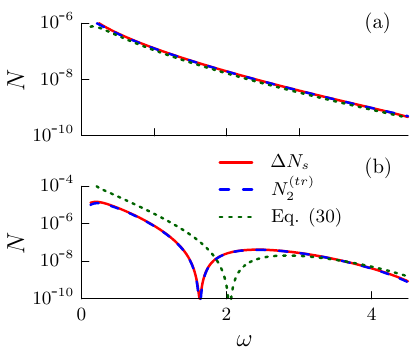}
  \caption{The change in the number of particles of the soliton
    $\Delta N_s$ (red solid lines) with $\gamma = -0.125$ (a) and
    with $\gamma = -0.1874375$ (b) and the number of particles in the
    $\psi_2^{(tr)}$ wave packet $N_2^{(tr)}$ (blue dashed lines) as a function
    of the incident wave packet frequency. The green dotted lines represent
    the estimation for the norm $\psi_2^{(tr)}$ derived from Eq.~\eqref{tr2}
    (see Sec.~\ref{pert}). The other parameters are the same as
    in Fig.~\ref{fig:evol}.}
  \label{fig:dnsol}
\end{figure}

The above argument indicates that the scattering of the wave packet
is accompanied by an increase in the number of particles in the soliton.
This is confirmed by numerical simulations. The change in the norm
of the soliton as a function of the incident wave packet frequency is
presented in Fig.~\ref{fig:dnsol} by solid line. Surprisingly that this
change coincides with high accuracy with the number of particles
in the transmitted wave packet $\psi_2^{(tr)}$ (dahsed line
in Fig~\ref{fig:dnsol}), 
\begin{equation}
  \label{NsN2}
  \Delta N_s = N_2^{(tr)}.
\end{equation}
Considering that $\psi_2^{(tr)} \propto A^2$ the number of particles
captured by the soliton is parametrically small, $\Delta N_s \propto A^4$. 

Substituting Eq.~\eqref{NsN2} into Eq.~\eqref{Nlaw} we obtain 
$\Delta N_1 = -2 \Delta N_s$. This equation together with Eq.~\eqref{dE} leads
to $\Delta E_s = \gamma \Delta N_s$ (cf Eq.~\eqref{dEsol}). This relation
implies that the particles captured by the soliton settle in the ground state
with energy $\gamma$, while the exited soliton modes remain unoccupied. 

Let us note that Eq.~\eqref{NsN2} is remarkable in the following respect.
Instead of calculating the change in the soliton's number of particles one
can find the norm of the wave packet $\psi_2^{(tr)}$ and use Eq.~\eqref{NsN2}.
The calculation of the norm of $\psi_2^{(tr)}$ in turn can be performed
analytically, as we will show in the next section.

\section{Perturbation theory}
\label{pert}
Now, we construct a perturbation theory based on the small
amplitude $A$ of a wave packet scattered on a soliton. First, we consider
$A \ll f(x)$. Second, for simplicity we will neglect the dependence of
$A$ on $x$, assuming that the width of the wave packet is much larger than
the size of the soliton. In this case, the wave packet can be considered
as a plane wave. Third, we assume that $|V(A^2)| \ll \omega$,
so the wave freely propagates to the region of interaction with the
soliton.

Thus, the solution of the equation of motion~\eqref{schrod} can be found
in the form of a series
\begin{equation}
  \label{perturb}
  \psi = \psi_s + \psi_1 + \psi_2 + \dots\;,
\end{equation}
where $\psi_n \propto A^n$. Consider each of these perturbations separately.

\subsection{First order}

Substituting Eq.~\eqref{perturb} into Eq.~\eqref{schrod} and combining the
terms of order of $A$ we arrive to the liner equation
\begin{equation}
  \label{lin1}
  i\partial_t \psi_1 + \frac{1}{2}\partial^2_x\psi_1
  = \left (V^{0} + V^{(1)} |\psi_s|^2 \right)  \psi_1
  + V^{(1)}\psi_s^2\psi_1^*\;,
\end{equation}
where we expand the potential in a Taylor series and introduce the notations 
\begin{equation}
  \label{der}
  \begin{split}
    &V^{(0)}(x) = V(|\psi_s|^2)\;,\\
    &V^{(i)}(x) = \left . \frac{d^i V(s)}{d s^i} \right |_{s=|\psi_s|^2}\;.
  \end{split}
\end{equation}
For the scattering problem under consideration the wave function $\psi_1$
satisfies the boundary condition~\eqref{psi0} at $x \to -\infty$ with
$A(t,x) = \mathrm{const}$. Eq.~\eqref{lin1} describes the incident wave
scattering on the soliton in the leading order. It can be solved
analytically for several specific models. In the general case 
$\psi_1$ can be found for $\omega \gg -\gamma$. In this limit the last term
in Eq.~\eqref{lin1} oscillates rapidly compared to the others, which can be
seen by multiplying the equation by $\mathrm{e}^{i\omega t}$, and can be
neglected. As a result, we arrive to the linear Sch\"odinger equation for
the wave function $\psi_1$ in the effective potential $V^{eff}_1(x)$, where
\begin{equation}
  \label{V1}
  V^{eff}_1(x) = V^{(0)} +V^{(1)}f^2\;.
\end{equation}
This equation can be solved by means of semiclassical approach, which is
applicable in the considered high-frequency limit. Introducing the classical
momentum,  
\begin{equation}
  \label{momentum}
  p_\omega(x) = \sqrt{2(\omega-V^{eff}_1(x))}\;,
\end{equation}
for the transmitted wave one writes
\begin{equation}
  \label{tr1}
  \begin{split}
    & \psi_1^{(tr)} = A_1^{(tr)}(x)\cdot \mathrm{e}^{-i\omega t + i S_\omega(x)}\;,\\
    & A_1^{(tr)}(x) = \frac{A\sqrt{p}}{\sqrt{p_\omega(x)}}\;,
  \end{split}
\end{equation}
where
\begin{equation}
  \label{action}
  S_{\omega}(x) = \int^x p_\omega(y) dy
\end{equation}
is the time-independent part of the classical action and 
the amplitude is normalized in accordance with Eq.~\eqref{psi0}.

\begin{figure}[t!]
  \centering
  \includegraphics{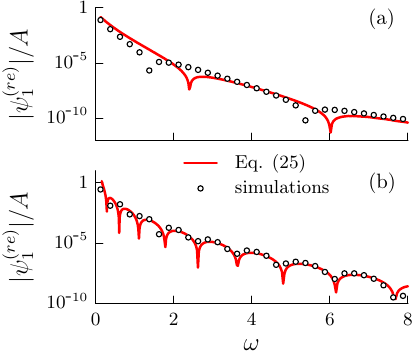}
  \caption{The amplitude of the left moving wave $\psi_1^{(re)}$ created
    in the scattering on the soliton with $\gamma = -0.125$ (a) and with
    $\gamma = -0.1874375$ (b) as a function of the incident wave
    frequency. Solid lines represent theoretical results,
    see Eq.~\eqref{ref1}, circles show the results of numerical simulations.}
  \label{fig:refl}
\end{figure}

The reflected wave also can be found by semiclassical method. However,
in this paper we apply another approach base on the Born approximation
following Ref.~\cite{Chan:2022bkz}, which
is simple and sufficiently accurate. Let us substitute
$\psi_1 = \psi_0 + \delta \psi_1$, where $\psi_0$ is a freely propagating
plane wave~\eqref{psi0}, into the linear Schr\"odinger equation. The solution
for $\delta \psi_1$ can be written as
\begin{equation}
  \label{corr}
  \delta \psi_1 = \int dt' dx' G(t-t',x-x')V^{eff}_1(x')\psi_0(t',x')\;, 
\end{equation}
where $G$ is the retarded Green's function. As we discussed above, the
evolution of a high-frequency waves in potential $V_1^{eff}$ proceeds
semiclassically. In this regime for the Green's function one writes
\begin{equation}
  \label{green}
  G(t,x,x') = \frac{1}{(2\pi)^2}\int d\Omega d k
  \frac{\psi_{\omega_k}(x)\psi^*_{\omega_k}(x')}{\Omega - \omega_k + i0}
  \mathrm{e}^{-i\Omega t}\;,
\end{equation}
where $\psi_{\omega_k}$ is the time-independent semiclassical wavefunction,
\begin{equation}
  \psi_{\omega_k}(x) = \frac{\sqrt{k}}{\sqrt{p_{\omega_k}(x)}}
  \mathrm{e}^{i S_{\omega_k}(x)}\;, \qquad
  \omega_k = \frac{k^2}{2}\;.
\end{equation}
Performing integration with respect to $t'$, $\Omega$ and $k$ in
Eq.~\eqref{corr} for the reflected wave we obtain
\begin{equation}
  \label{ref1}
  \begin{split}
  &\psi_1^{(re)} = A_1^{(re)}(x)\cdot\mathrm{e}^{-i\omega t - iS_\omega(x)}\;,\\
  &A_1^{(re)}(x) = \frac{-iA}{\sqrt{p_\omega(x)}}\int_x^{\infty} dx'
  \frac{V^{eff}_1(x')\, \mathrm{e}^{ipx'+iS_\omega(x')}}{\sqrt{p_\omega(x')}}\;.
  \end{split}
\end{equation}
Note, that the integrand in Eq.~\eqref{ref1} oscillates rapidly with
changing $x'$ so that the amplitude of the reflected wave is expected to be
exponentially small\footnote{The integration in this case can be done by the
  stationary phase method.}. 

In Fig.~\ref{fig:refl} we compare predictions of Eq.~\eqref{ref1} for the
ampitude of $\psi_1^{(re)}$ at large negative $x$ with numerical results in
the model~\eqref{pot} presented in Sec.~\ref{numerics}. For this purpose the
integral in Eq.~\eqref{ref1} was computed numerically. One observes that the
analytic predictions (lines in the figure) coincide with the numerical results
(circles in the figure) in the high frequency limit $\omega \gg -\gamma$ but
deviate from them at small $\omega$. 

\subsection{Second order}

Now let us consider the scattering of the wave on the solitons in the second
order in its amplitude. Substituting Eq.~\eqref{perturb} into
Eq.~\eqref{schrod} and collecting the term of order of $A^2$ we obtain 
\begin{equation}
  \label{lin2}
   \begin{split}
   & i\partial_t \psi_2 + \frac{1}{2}\partial^2_x\psi_2
  -V^{eff}_1 \psi_2 - V^{(1)} \psi_s^2 \psi^*_2 \\
  & = \left (V^{(1)} + \frac12 V^{(2)}|\psi_s|^2 \right ) \psi_1^2 \psi_s^*\\
  & + \left ( 2 V^{(1)} + V^{(2)}|\psi_s|^2 \right ) |\psi_1|^2 \psi_s\\
  & + \frac12 V^{(2)} (\psi_1^*)^2\psi_s^3\;.
   \end{split}
\end{equation}
Due to the source in the right-hand side, this equation describes the
creation of a second-order wave function $\psi_2$ as a result of the
interaction of an incident wave $\psi_1$ with a soliton. There are three
terms in the source that oscillate at different frequencies. Each term
generates its own distinct contribution to $\psi_2$ and can be analyzed
separately.

\begin{figure}[t!]
  \centering
  \includegraphics{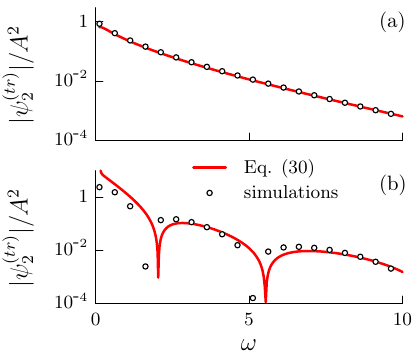}
  \caption{The amplitude of the right moving wave $\psi_2^{(tr)}$ created
    by nonlinear interaction between the soliton with $\gamma = -0.125$ (a) and
    with $\gamma = -0.1874375$ (b) and the incident wave as a function of its
    frequency. Solid lines represent theoretical results,
    see Eq.~\eqref{tr2}, circles show the results of numerical simulations.}
  \label{fig:tr2}
\end{figure}

The time-dependence of the first term of the source in Eq.~\eqref{lin2} is
given by $\psi_1^2 \psi_s^*$ multiplier. As a result it supports creation of
the wave $\psi_2$ with frequency $\omega_2 = 2 \omega - \gamma$ propagating
from the soliton. This process can be described analitically in
high-frequency regime $\omega \gg - \gamma$. In this case the last term
in the left-hand side of Eq.~\eqref{lin2} is highly-oscillating thus
can be neglected. As a result, we obtain the linear Schr\"odinger equation
\begin{equation}
  \label{schrod2}
  i\partial_t \psi_2 + \frac{1}{2}\partial^2_x\psi_2 -V^{eff}_1 \psi_2
   = V^{eff}_2 \psi_1^2 \mathrm{e}^{i\gamma t}\;,
\end{equation}
where we introduce
\begin{equation}
  V^{eff}_2(x) = V^{(1)}f + \frac12 V^{(2)}f^3\;.
\end{equation}
Its solution for high-frequency waves can be obtained by using semiclassical
Green's function~\eqref{green}, 
\begin{equation}
  \label{psi2}
  \psi_2 = \int dt' dx' G(t-t',x,x')V^{eff}_2(x')
  \psi^2_1(t',x')\mathrm{e}^{i \gamma t'}\;.
\end{equation}
Here $\psi_1 = \psi_1^{(tr)} + \psi_1^{(re)}$. For the wave propagating to the
right the main contribution comes from the transmitted
wave~\eqref{tr1}. Integrating over $t'$, $\Omega$ and $k$ we get
\begin{equation}
  \label{tr2}
  \begin{split}
    &\psi_2^{(tr)}=A_2^{tr}(x)\cdot\mathrm{e}^{-i\omega_2t + iS_{\omega_2}(x)}\;,\\
    &A_2^{tr}(x) = \frac{-i A^2\,p} {\sqrt{p_{\omega_2}(x)}}
    \int_{-\infty}^x dx'\,\frac{V_2^{eff}(x')\,
      \mathrm{e}^{2iS_\omega(x') - iS_{\omega_2}(x')}}
        {p_\omega(x')\sqrt{p_{\omega_2}(x')}}\;.
  \end{split}
\end{equation}

Similarly, from Eq.~\eqref{psi2} for reflected wave we get
\begin{equation}
  \label{re2}
  \begin{split}
    &\psi_2^{(re)}=A_2^{re}(x)\cdot\mathrm{e}^{-i\omega_2t - iS_{\omega_2}(x)}\;,\\
    &A_2^{re}(x) = \frac{i A^2} {\sqrt{p_{\omega_2}(x)}}
    \int_x^{\infty} dx'\,V_2^{eff}(x')\,
    \frac{\mathrm{e}^{iS_{\omega_2}(x')}}{\sqrt{p_{\omega_2}(x')}}\\
    &\times
    \left (A_1^{(tr)}(x')\mathrm{e}^{iS_\omega(x')} +
    A_1^{(re)}(x')\mathrm{e}^{-iS_\omega(x')}\right)^2\;,
  \end{split}
\end{equation}
where the contributions from the transmitted and reflected waves of the first
order, given by Eqs.~\eqref{tr1},~\eqref{ref1}, have been taken into account.

\begin{figure}[t!]
  \centering
  \includegraphics{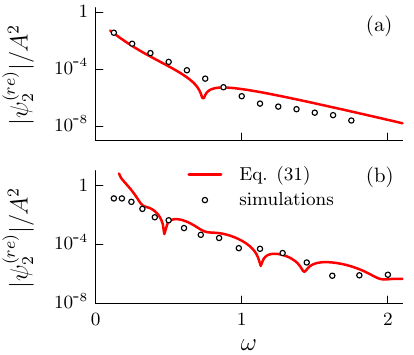}
  \caption{The amplitude of the left moving wave $\psi_2^{(re)}$ created
    by nonlinear interaction between the soliton with $\gamma = -0.125$ (a) and
    with $\gamma = -0.1874375$ (b) and the incident wave as a function of its
    frequency. Solid lines represent theoretical results,
    see Eq.~\eqref{re2}, circles show the results of numerical simulations.}
  \label{fig:re2}
\end{figure}

The amplitudes of $\psi_2^{(tr)}$ and $\psi_2^{(re)}$ obtained by numerical
integration in Eqs.~\eqref{tr2},~\eqref{re2} for the model~\eqref{pot} in
comparison with the results of numerical simulations are presented in
Fig.~\ref{fig:tr2} and Fig.~\ref{fig:re2} respectively. As one can see
from the figures, the analytical results are in good agreement with the
numerical data.

As we discussed in the previous section, the norm of $\psi_2^{(tr)}$ equals the
number of particles absorbed by the soliton. As one can see from
Fig.~\ref{fig:evol}, the width of the wave packets leaving the interaction
region after scattering on the soliton coincides with the width of the
incident wave packet. Thus, $\psi_2^{(tr)}$ is a Gaussian wave packet, see
Eq.~\eqref{gauss}, with the amplitude estimated by Eq.~\eqref{tr2}.
Calculating its norm, we obtain an estimate for the number of particles
absorbed by the soliton, which is shown (by green dotted lines) in
Fig.~\ref{fig:dnsol}.

Now let us discuss the second and third terms of the source in the
right-hand side of Eq.~\eqref{lin2}. These terms oscillate with negative
frequencies, $\gamma$ and $3\gamma-2\omega$ respectively, and may
contribute to the emergence of localized modes associated with changes
in the number of particles in the soliton. However, the analytical search
for non-propagating solutions is very challenging. In this case the last term
of the operator in the left-hand side of Eq.~\eqref{lin2} cannot be neglected
and its inversion appears to be rather complicated
(see, however, \cite{Kovtun:2021rcm}). Given that the change in the number of
particles in the soliton $\propto A^4$, following this approach we will have to
solve equations up to the fourth order of perturbation theory. In this way,
the method outlined above for calculation of the soliton norm change is simpler
and more convenient.

\section{Discussion}
\label{discus}

In this study, we have demonstrated that wave scattering by a soliton is
accompanied by the generation of waves with frequencies that are multiples
of the incident wave frequency, minus the soliton frequency. Additionally, we
have observed that the soliton undergoes growth by absorbing particles from
the incident wave into its ground state, while the exited states remain
unoccupied. This fact can be understood by adopting the methods of quantum
field theory\footnote{It is important to emphasize that quantum physics is
  not essential for the soliton-wave interaction}. The nonlinear
self-interaction term $|\psi|^4$ is responsible for scattering of two
particles from the incident wave packet into two other states. The occupation
number of the soliton ground state is large, thus the transition to this
state is strongly Bose-enhanced. Based on the Born approximation, a wave
packet with a frequency of $2 \omega - \gamma$ is also produced in this
process. According to the conservation laws of total energy and total number
of particles, it follows that the norm of this wave packet should be equal
to the number of particles absorbed by the soliton.

The scattering of a wide, low-amplitude soliton on a high-amplitude soliton
proceeds similarly. We have verified that in this case a reflected wave packet
is produced, as well as wave packets with frequencies that are integer
multiples of the incident soliton frequency, minus the frequency of the
stationary soliton. The shape of the wave packets precisely matches the shape
of the incident soliton. Their amplitudes can be estimated by the method
developed in Section~\ref{pert}. 

It should also be noted, that we expect similar processes in the scattering of a plane wave on a soliton in 3+1 dimensions, and our analytical calculation method can be easily generalized to this case.

\acknowledgments
We thank M. Smolyakov and E. Nugaev for numerous discussions and interest in the work. The work was supported by the grant RSF 22-12-00215 and, in its numerical
part, by the “BASIS” foundation. Numerical calculations were performed on the
Computational Cluster of the Theoretical Division of~INR RAS.

\appendix
\section{Numerical scheme}
\label{numer}

We solve numerically the nonlinear Schr\"odinger equation~\eqref{schrod} using
4-th order pseudo-spectral operator-splitting method~\cite{YOSHIDA1990262,
  McLachlan}. This method is unitary, stable, symplectic and therefore
exceptionally accurate for computing small-valued high-order effects. The wave
function at the next time step is given by
\begin{equation}
  \label{evolution}
  \psi(t+\Delta t) = \hat{\cal U}\, \psi(t)\;,
\end{equation}
where the evolution operator is replaced by discrete formula 
\begin{equation}
  \label{U}
  \hat{\cal U} = \prod_{\alpha = 1}^{4} \mathrm{e}^{-i d_\alpha \Delta t\, V_\alpha(x)}
  \mathrm{e}^{-i c_\alpha \Delta t\, \hat{p}^2/2}\; + O(\Delta t^5).
\end{equation}
Here the product is ordered right-to-left, its parameters $c_\alpha$ and
$d_\alpha$ are given in~\cite{YOSHIDA1990262, McLachlan}, and the "potentials"
$V_\beta$ are computed using the “current” field $\psi(t)$ (multiplied by all
operators with $\alpha \leq \beta$), i.e. $V_\beta = V(|\psi|^2)$.
Eq.~\eqref{U} breaks the time interval $\Delta t$ into $c$- and
$d$-sub-intervals, on which the "kinetic" part and the "potential" part of
the evolution operator are used respectively.

The numerical application of the formula~\eqref{U} is as follows. We introduce 
uniform lattice with $N$ sites in the finite box of size $L$. The field
values $\psi(t,x_j)$ are stored on the lattices sites
$x_j = -L/2 + j \Delta x$, $0 \leq j < N$, where $\Delta x = L/N$. Typically,
we use $L = 12000$ with $N = 2^{15}$ and switch to another values for
resolution tests. 

Time evolution of $\psi(x_j)$ is calculated by sequentially acting with
operators in Eq.~\eqref{U}. The action of the "kinetic" parts of the
evolution operator $\mathrm{e}^{-i c_\alpha \Delta t\, \hat{p}^2/2}$ is performed
as follows. First, we Fourier-transform the field,
$\psi(x_j) = \sum_k \tilde{\psi}(p_k)\mathrm{e}^{-ip_kx_j}$, using Fast Fourier transform algorithm, where
$\tilde{\psi}(p_k)$ is the image at discrete momenta $p_k = 2\pi k/L$, 
$-N/2 < k \leq N/2$. After that multiplication by "kinetic" part of the
evolution operator corresponds to phase rotation $\tilde{\psi}(p_k) \to
\tilde{\psi}(p_k)\, \mathrm{e}^{-i c_1 \Delta t\, p_k^2/2}$. Then we return
to the coordinate representation for the $\psi$ with the inverse Fourier
transform and multimply it by the "potential" part of the evolution operator. 
These actions are repeated the required number of times according to
Eq.~\eqref{U}.

Each step of the algorithm is reduced to multiplying the wave function
by a phase, so it exectly (up to truncation errors) preserves the norm. The
total energy is not conserved. However, we have chosen a time step such
that the relative non-conservation of energy in our solutions has never
exceeded $10^{-12}$. This accuracy allows us to observe the appearance of
wave packets with an amplitude up to $10^{-10}$.

\bibliography{literature}

\begin{thebibliography}{22}%
\makeatletter
\providecommand \@ifxundefined [1]{%
 \@ifx{#1\undefined}
}%
\providecommand \@ifnum [1]{%
 \ifnum #1\expandafter \@firstoftwo
 \else \expandafter \@secondoftwo
 \fi
}%
\providecommand \@ifx [1]{%
 \ifx #1\expandafter \@firstoftwo
 \else \expandafter \@secondoftwo
 \fi
}%
\providecommand \natexlab [1]{#1}%
\providecommand \enquote  [1]{``#1''}%
\providecommand \bibnamefont  [1]{#1}%
\providecommand \bibfnamefont [1]{#1}%
\providecommand \citenamefont [1]{#1}%
\providecommand \href@noop [0]{\@secondoftwo}%
\providecommand \href [0]{\begingroup \@sanitize@url \@href}%
\providecommand \@href[1]{\@@startlink{#1}\@@href}%
\providecommand \@@href[1]{\endgroup#1\@@endlink}%
\providecommand \@sanitize@url [0]{\catcode `\\12\catcode `\$12\catcode
  `\&12\catcode `\#12\catcode `\^12\catcode `\_12\catcode `\%12\relax}%
\providecommand \@@startlink[1]{}%
\providecommand \@@endlink[0]{}%
\providecommand \url  [0]{\begingroup\@sanitize@url \@url }%
\providecommand \@url [1]{\endgroup\@href {#1}{\urlprefix }}%
\providecommand \urlprefix  [0]{URL }%
\providecommand \Eprint [0]{\href }%
\providecommand \doibase [0]{http://dx.doi.org/}%
\providecommand \selectlanguage [0]{\@gobble}%
\providecommand \bibinfo  [0]{\@secondoftwo}%
\providecommand \bibfield  [0]{\@secondoftwo}%
\providecommand \translation [1]{[#1]}%
\providecommand \BibitemOpen [0]{}%
\providecommand \bibitemStop [0]{}%
\providecommand \bibitemNoStop [0]{.\EOS\space}%
\providecommand \EOS [0]{\spacefactor3000\relax}%
\providecommand \BibitemShut  [1]{\csname bibitem#1\endcsname}%
\let\auto@bib@innerbib\@empty
\bibitem [{\citenamefont {Russell}(1844)}]{russell1844report}%
  \BibitemOpen
  \bibfield  {author} {\bibinfo {author} {\bibfnamefont {J.~S.}\ \bibnamefont
  {Russell}},\ }in\ \href@noop {} {\emph {\bibinfo {booktitle} {14th meeting of
  the British Association for the Advancement of Science}}},\ Vol.\ \bibinfo
  {volume} {311}\ (\bibinfo {year} {1844})\ p.\ \bibinfo {pages}
  {1844}\BibitemShut {NoStop}%
\bibitem [{\citenamefont {Kivshar}\ and\ \citenamefont
  {Agrawal}(2003)}]{kivshar2003optical}%
  \BibitemOpen
  \bibfield  {author} {\bibinfo {author} {\bibfnamefont {Y.}~\bibnamefont
  {Kivshar}}\ and\ \bibinfo {author} {\bibfnamefont {G.}~\bibnamefont
  {Agrawal}},\ }\href {https://books.google.ru/books?id=zzWgibj4ypsC} {\emph
  {\bibinfo {title} {Optical Solitons: From Fibers to Photonic Crystals}}}\
  (\bibinfo  {publisher} {Elsevier Science},\ \bibinfo {year}
  {2003})\BibitemShut {NoStop}%
\bibitem [{\citenamefont {Ichikawa}(1979)}]{Ichikawa_1979}%
  \BibitemOpen
  \bibfield  {author} {\bibinfo {author} {\bibfnamefont {Y.~H.}\ \bibnamefont
  {Ichikawa}},\ }\href {\doibase 10.1088/0031-8949/20/3-4/002} {\bibfield
  {journal} {\bibinfo  {journal} {Physica Scripta}\ }\textbf {\bibinfo {volume}
  {20}},\ \bibinfo {pages} {296} (\bibinfo {year} {1979})}\BibitemShut
  {NoStop}%
\bibitem [{\citenamefont {Bishop}\ \emph {et~al.}(1980)\citenamefont {Bishop},
  \citenamefont {Krumhansl},\ and\ \citenamefont {Trullinger}}]{BISHOP19801}%
  \BibitemOpen
  \bibfield  {author} {\bibinfo {author} {\bibfnamefont {A.}~\bibnamefont
  {Bishop}}, \bibinfo {author} {\bibfnamefont {J.}~\bibnamefont {Krumhansl}}, \
  and\ \bibinfo {author} {\bibfnamefont {S.}~\bibnamefont {Trullinger}},\
  }\href {\doibase https://doi.org/10.1016/0167-2789(80)90003-2} {\bibfield
  {journal} {\bibinfo  {journal} {Physica D: Nonlinear Phenomena}\ }\textbf
  {\bibinfo {volume} {1}},\ \bibinfo {pages} {1} (\bibinfo {year}
  {1980})}\BibitemShut {NoStop}%
\bibitem [{\citenamefont {Kusenko}\ and\ \citenamefont
  {Shaposhnikov}(1998)}]{KUSENKO199846}%
  \BibitemOpen
  \bibfield  {author} {\bibinfo {author} {\bibfnamefont {A.}~\bibnamefont
  {Kusenko}}\ and\ \bibinfo {author} {\bibfnamefont {M.}~\bibnamefont
  {Shaposhnikov}},\ }\href {\doibase
  https://doi.org/10.1016/S0370-2693(97)01375-0} {\bibfield  {journal}
  {\bibinfo  {journal} {Physics Letters B}\ }\textbf {\bibinfo {volume}
  {418}},\ \bibinfo {pages} {46} (\bibinfo {year} {1998})}\BibitemShut
  {NoStop}%
\bibitem [{\citenamefont {Lax}(1968)}]{Lax1968}%
  \BibitemOpen
  \bibfield  {author} {\bibinfo {author} {\bibfnamefont {P.~D.}\ \bibnamefont
  {Lax}},\ }\href {\doibase https://doi.org/10.1002/cpa.3160210503} {\bibfield
  {journal} {\bibinfo  {journal} {Communications on Pure and Applied
  Mathematics}\ }\textbf {\bibinfo {volume} {21}},\ \bibinfo {pages} {467}
  (\bibinfo {year} {1968})}\BibitemShut {NoStop}%
\bibitem [{\citenamefont {{Zakharov}}\ and\ \citenamefont
  {{Shabat}}(1972)}]{Zakharov1971}%
  \BibitemOpen
  \bibfield  {author} {\bibinfo {author} {\bibfnamefont {V.~E.}\ \bibnamefont
  {{Zakharov}}}\ and\ \bibinfo {author} {\bibfnamefont {A.~B.}\ \bibnamefont
  {{Shabat}}},\ }\href
  {http://www.jetp.ras.ru/cgi-bin/e/index/e/34/1/p62?a=list} {\bibfield
  {journal} {\bibinfo  {journal} {Sov. Phys. JETP}\ }\textbf {\bibinfo {volume}
  {34}},\ \bibinfo {pages} {62} (\bibinfo {year} {1972})}\BibitemShut {NoStop}%
\bibitem [{\citenamefont {Novikov}\ \emph {et~al.}(1984)\citenamefont
  {Novikov}, \citenamefont {Manakov}, \citenamefont {Pitaevskii},\ and\
  \citenamefont {Zakharov}}]{novikov1984theory}%
  \BibitemOpen
  \bibfield  {author} {\bibinfo {author} {\bibfnamefont {S.}~\bibnamefont
  {Novikov}}, \bibinfo {author} {\bibfnamefont {S.~V.}\ \bibnamefont
  {Manakov}}, \bibinfo {author} {\bibfnamefont {L.~P.}\ \bibnamefont
  {Pitaevskii}}, \ and\ \bibinfo {author} {\bibfnamefont {V.~E.}\ \bibnamefont
  {Zakharov}},\ }\href@noop {} {\emph {\bibinfo {title} {Theory of solitons:
  the inverse scattering method}}}\ (\bibinfo  {publisher} {Springer Science \&
  Business Media},\ \bibinfo {year} {1984})\BibitemShut {NoStop}%
\bibitem [{\citenamefont {Drazin}\ and\ \citenamefont
  {Johnson}(1989)}]{drazin1989solitons}%
  \BibitemOpen
  \bibfield  {author} {\bibinfo {author} {\bibfnamefont {P.~G.}\ \bibnamefont
  {Drazin}}\ and\ \bibinfo {author} {\bibfnamefont {R.~S.}\ \bibnamefont
  {Johnson}},\ }\href@noop {} {\emph {\bibinfo {title} {Solitons: an
  introduction}}},\ Vol.~\bibinfo {volume} {2}\ (\bibinfo  {publisher}
  {Cambridge university press},\ \bibinfo {year} {1989})\BibitemShut {NoStop}%
\bibitem [{\citenamefont {Zakharov}\ and\ \citenamefont
  {Shabat}(1973)}]{zakharov1973interaction}%
  \BibitemOpen
  \bibfield  {author} {\bibinfo {author} {\bibfnamefont {V.~E.}\ \bibnamefont
  {Zakharov}}\ and\ \bibinfo {author} {\bibfnamefont {A.~B.}\ \bibnamefont
  {Shabat}},\ }\href@noop {} {\bibfield  {journal} {\bibinfo  {journal} {Sov.
  Phys. JETP}\ }\textbf {\bibinfo {volume} {37}},\ \bibinfo {pages} {823}
  (\bibinfo {year} {1973})}\BibitemShut {NoStop}%
\bibitem [{\citenamefont {Kivshar}\ and\ \citenamefont
  {Malomed}(1989)}]{Kishvar1989}%
  \BibitemOpen
  \bibfield  {author} {\bibinfo {author} {\bibfnamefont {Y.~S.}\ \bibnamefont
  {Kivshar}}\ and\ \bibinfo {author} {\bibfnamefont {B.~A.}\ \bibnamefont
  {Malomed}},\ }\href {\doibase 10.1103/RevModPhys.61.763} {\bibfield
  {journal} {\bibinfo  {journal} {Rev. Mod. Phys.}\ }\textbf {\bibinfo {volume}
  {61}},\ \bibinfo {pages} {763} (\bibinfo {year} {1989})}\BibitemShut
  {NoStop}%
\bibitem [{\citenamefont {Gatz}\ and\ \citenamefont
  {Herrmann}(1992)}]{Gatz1992}%
  \BibitemOpen
  \bibfield  {author} {\bibinfo {author} {\bibfnamefont {S.}~\bibnamefont
  {Gatz}}\ and\ \bibinfo {author} {\bibfnamefont {J.}~\bibnamefont
  {Herrmann}},\ }\href {\doibase 10.1109/3.142561} {\bibfield  {journal}
  {\bibinfo  {journal} {IEEE Journal of Quantum Electronics}\ }\textbf
  {\bibinfo {volume} {28}},\ \bibinfo {pages} {1732} (\bibinfo {year}
  {1992})}\BibitemShut {NoStop}%
\bibitem [{\citenamefont {Tikhonenko}\ \emph {et~al.}(1996)\citenamefont
  {Tikhonenko}, \citenamefont {Christou},\ and\ \citenamefont
  {Luther-Davies}}]{Tikhonenko1996}%
  \BibitemOpen
  \bibfield  {author} {\bibinfo {author} {\bibfnamefont {V.}~\bibnamefont
  {Tikhonenko}}, \bibinfo {author} {\bibfnamefont {J.}~\bibnamefont
  {Christou}}, \ and\ \bibinfo {author} {\bibfnamefont {B.}~\bibnamefont
  {Luther-Davies}},\ }\href {\doibase 10.1103/PhysRevLett.76.2698} {\bibfield
  {journal} {\bibinfo  {journal} {Phys. Rev. Lett.}\ }\textbf {\bibinfo
  {volume} {76}},\ \bibinfo {pages} {2698} (\bibinfo {year}
  {1996})}\BibitemShut {NoStop}%
\bibitem [{\citenamefont {Nguyen}\ \emph {et~al.}(2014)\citenamefont {Nguyen},
  \citenamefont {Dyke}, \citenamefont {Luo}, \citenamefont {Malomed},\ and\
  \citenamefont {Hulet}}]{Nguyen2014}%
  \BibitemOpen
  \bibfield  {author} {\bibinfo {author} {\bibfnamefont {J.~H.~V.}\
  \bibnamefont {Nguyen}}, \bibinfo {author} {\bibfnamefont {P.}~\bibnamefont
  {Dyke}}, \bibinfo {author} {\bibfnamefont {D.}~\bibnamefont {Luo}}, \bibinfo
  {author} {\bibfnamefont {B.~A.}\ \bibnamefont {Malomed}}, \ and\ \bibinfo
  {author} {\bibfnamefont {R.~G.}\ \bibnamefont {Hulet}},\ }\href {\doibase
  10.1038/nphys3135} {\bibfield  {journal} {\bibinfo  {journal} {Nature
  Physics}\ }\textbf {\bibinfo {volume} {10}},\ \bibinfo {pages} {918}
  (\bibinfo {year} {2014})}\BibitemShut {NoStop}%
\bibitem [{\citenamefont {Garnier}\ and\ \citenamefont
  {Abdullaev}(2006)}]{garnier2006transmission}%
  \BibitemOpen
  \bibfield  {author} {\bibinfo {author} {\bibfnamefont {J.}~\bibnamefont
  {Garnier}}\ and\ \bibinfo {author} {\bibfnamefont {F.~K.}\ \bibnamefont
  {Abdullaev}},\ }\href@noop {} {\bibfield  {journal} {\bibinfo  {journal}
  {Physical Review A}\ }\textbf {\bibinfo {volume} {74}},\ \bibinfo {pages}
  {013604} (\bibinfo {year} {2006})}\BibitemShut {NoStop}%
\bibitem [{\citenamefont {Haus}\ \emph {et~al.}(1996)\citenamefont {Haus},
  \citenamefont {Khatri}, \citenamefont {Wong}, \citenamefont {Ippen},\ and\
  \citenamefont {Tamura}}]{haus1996interaction}%
  \BibitemOpen
  \bibfield  {author} {\bibinfo {author} {\bibfnamefont {H.~A.}\ \bibnamefont
  {Haus}}, \bibinfo {author} {\bibfnamefont {F.~I.}\ \bibnamefont {Khatri}},
  \bibinfo {author} {\bibfnamefont {W.~S.}\ \bibnamefont {Wong}}, \bibinfo
  {author} {\bibfnamefont {E.~P.}\ \bibnamefont {Ippen}}, \ and\ \bibinfo
  {author} {\bibfnamefont {K.~R.}\ \bibnamefont {Tamura}},\ }\href@noop {}
  {\bibfield  {journal} {\bibinfo  {journal} {IEEE journal of quantum
  electronics}\ }\textbf {\bibinfo {volume} {32}},\ \bibinfo {pages} {917}
  (\bibinfo {year} {1996})}\BibitemShut {NoStop}%
\bibitem [{\citenamefont {Vakhitov}\ and\ \citenamefont
  {Kolokolov}(1973)}]{vakhitov1973}%
  \BibitemOpen
  \bibfield  {author} {\bibinfo {author} {\bibfnamefont {N.}~\bibnamefont
  {Vakhitov}}\ and\ \bibinfo {author} {\bibfnamefont {A.}~\bibnamefont
  {Kolokolov}},\ }\href@noop {} {\bibfield  {journal} {\bibinfo  {journal}
  {Radiophysics and Quantum Electronics}\ }\textbf {\bibinfo {volume} {16}},\
  \bibinfo {pages} {783} (\bibinfo {year} {1973})}\BibitemShut {NoStop}%
\bibitem [{\citenamefont {Dmitriev}\ \emph {et~al.}(2023)\citenamefont
  {Dmitriev}, \citenamefont {Dmitrieva},\ and\ \citenamefont {Panin}}]{movie}%
  \BibitemOpen
  \bibfield  {author} {\bibinfo {author} {\bibfnamefont {A.}~\bibnamefont
  {Dmitriev}}, \bibinfo {author} {\bibfnamefont {E.}~\bibnamefont {Dmitrieva}},
  \ and\ \bibinfo {author} {\bibfnamefont {A.}~\bibnamefont {Panin}},\
  }\href@noop {} {\enquote {\bibinfo {title} {Scattering of wave packet on a
  soliton},}\ }\bibinfo {howpublished} {\url{https://youtu.be/XYQlzUEoTfU}}
  (\bibinfo {year} {2023})\BibitemShut {NoStop}%
\bibitem [{\citenamefont {Chan}\ \emph {et~al.}(2022)\citenamefont {Chan},
  \citenamefont {Sibiryakov},\ and\ \citenamefont {Xue}}]{Chan:2022bkz}%
  \BibitemOpen
  \bibfield  {author} {\bibinfo {author} {\bibfnamefont {J.~H.-H.}\
  \bibnamefont {Chan}}, \bibinfo {author} {\bibfnamefont {S.}~\bibnamefont
  {Sibiryakov}}, \ and\ \bibinfo {author} {\bibfnamefont {W.}~\bibnamefont
  {Xue}},\ }\href@noop {} {\  (\bibinfo {year} {2022})},\ \Eprint
  {http://arxiv.org/abs/2207.04057} {arXiv:2207.04057 [astro-ph.CO]}
  \BibitemShut {NoStop}%
\bibitem [{\citenamefont {Kovtun}(2022)}]{Kovtun:2021rcm}%
  \BibitemOpen
  \bibfield  {author} {\bibinfo {author} {\bibfnamefont {A.}~\bibnamefont
  {Kovtun}},\ }\href {\doibase 10.1103/PhysRevD.105.036011} {\bibfield
  {journal} {\bibinfo  {journal} {Phys. Rev. D}\ }\textbf {\bibinfo {volume}
  {105}},\ \bibinfo {pages} {036011} (\bibinfo {year} {2022})},\ \Eprint
  {http://arxiv.org/abs/2110.05222} {arXiv:2110.05222 [hep-th]} \BibitemShut
  {NoStop}%
\bibitem [{\citenamefont {Yoshida}(1990)}]{YOSHIDA1990262}%
  \BibitemOpen
  \bibfield  {author} {\bibinfo {author} {\bibfnamefont {H.}~\bibnamefont
  {Yoshida}},\ }\href {\doibase https://doi.org/10.1016/0375-9601(90)90092-3}
  {\bibfield  {journal} {\bibinfo  {journal} {Physics Letters A}\ }\textbf
  {\bibinfo {volume} {150}},\ \bibinfo {pages} {262} (\bibinfo {year}
  {1990})}\BibitemShut {NoStop}%
\bibitem [{\citenamefont {McLachlan}(1993)}]{McLachlan}%
  \BibitemOpen
  \bibfield  {author} {\bibinfo {author} {\bibfnamefont {R.}~\bibnamefont
  {McLachlan}},\ }\href {\doibase 10.1007/BF01385708} {\bibfield  {journal}
  {\bibinfo  {journal} {Numerische Mathematik}\ }\textbf {\bibinfo {volume}
  {66}},\ \bibinfo {pages} {pages 465–492} (\bibinfo {year}
  {1993})}\BibitemShut {NoStop}%
\end{thebibliography}%

\end{document}